\documentclass[floatfix,twocolumn,aps,prl,reprint,superscriptaddress]{revtex4-1}
\usepackage[T1]{fontenc}
\usepackage{tikz}
\pdfoutput=1
\newcommand{\be}{\begin{equation}}
\newcommand{\ee}{\end{equation}}
\newcommand{\bea}{\begin{eqnarray}}
\newcommand{\eea}{\end{eqnarray}}
\newcommand{\bean}{\begin{eqnarray*}}
\newcommand{\eean}{\end{eqnarray*}}
\newcommand{\bra}[1]{\left\langle #1\right|}
\newcommand{\ket}[1]{\left|#1\right\rangle}

\newcommand{\ketbra}[2]{|#1\rangle\langle #2|}
\newcommand{\re}{\text{Re}}
\newcommand{\im}{\text{Im}}

\usepackage{xcolor}
\usepackage{cancel}
\usepackage{amsmath}
\DeclareMathOperator*{\minimize}{minimize}
\usepackage{tikz}
\usepackage{tikz-timing}
\usetikzlibrary{decorations.pathreplacing,angles,quotes}
\usetikzlibrary{arrows.meta,decorations.pathmorphing,decorations.pathreplacing,positioning,shapes}
\usepackage{dcolumn}
\usepackage[labelformat=simple]{subcaption}
\usepackage{bbold}
\usepackage{graphicx}
\usepackage{dcolumn}
\usepackage[font=small,labelfont=bf,figurename=Figure,tablename=Table]{caption}
\usepackage{siunitx}
\usepackage{relsize}
\usepackage{hyperref}

\usepackage[singlelinecheck=off,justification=raggedright]{subcaption}
\DeclareCaptionLabelSeparator{bar}{ | }
\usepackage{color}
\mathchardef\mhyphen="2D

\renewcommand\bra[1]{{\langle{#1}|}}
\makeatletter
\newcommand\k@t[1]{{|{#1}\rangle}}
\makeatother
\usepackage{color, colortbl}
\definecolor{Gray}{gray}{0.98}
\definecolor{storagecolor}{rgb}{1,0.97,0.9}
\definecolor{tmoncolor}{rgb}{0.92,1,1}
\definecolor{readoutcolor}{rgb}{1,0.9,0.9}

\begin{document}

\bibliographystyle{naturemag}

\title{
Efficient multimode Wigner tomography
}

\author{Kevin He}
\affiliation{James Franck Institute, University of Chicago, Chicago, Illinois 60637, USA}
\affiliation{Department of Physics, University of Chicago, Chicago, Illinois 60637, USA}

\author{Ming Yuan}
\affiliation{Pritzker School of Molecular Engineering, University of Chicago, Chicago, Illinois 60637, USA}

\author{Yat Wong}
\affiliation{Pritzker School of Molecular Engineering, University of Chicago, Chicago, Illinois 60637, USA}

\author{Srivatsan Chakram}
\affiliation{Department of Physics and Astronomy, Rutgers University, Piscataway, NJ 08854, USA}

\author{Alireza Seif}
\affiliation{Pritzker School of Molecular Engineering, University of Chicago, Chicago, Illinois 60637, USA}

\author{Liang Jiang}
\affiliation{Pritzker School of Molecular Engineering, University of Chicago, Chicago, Illinois 60637, USA}

\author{David I. Schuster}
\affiliation{James Franck Institute, University of Chicago, Chicago, Illinois 60637, USA}
\affiliation{Department of Physics, University of Chicago, Chicago, Illinois 60637, USA}
\affiliation{Pritzker School of Molecular Engineering, University of Chicago, Chicago, Illinois 60637, USA}

% \begin{refsection}
% \begin{bibunit}

\begin{abstract}
\noindent
Advancements in quantum system lifetimes and control have enabled the creation of increasingly complex quantum states, such as those on multiple bosonic cavity modes.
When characterizing these states, traditional tomography scales exponentially in both computational and experimental measurement requirement, which becomes prohibitive as the state size increases.
Here, we implement a state reconstruction method whose sampling requirement instead scales polynomially with subspace size, and thus mode number, for states that can be expressed within such a subspace. We demonstrate this improved scaling with Wigner tomography of multimode entangled W states of up to 4 modes on a 3D circuit quantum electrodynamics (cQED) system. This approach performs similarly in efficiency to existing matrix inversion methods for 2 modes, and demonstrates a noticeable improvement for 3 and 4 modes, with even greater theoretical gains at higher mode numbers.
\end{abstract}

\maketitle

Quantum state tomography (QST) is the process of determining the quantum state of a system, and is a fundamental part of quantum information processing. 
The exponentially greater complexity of quantum systems compared to classical ones makes efficient QST a challenging task. In its conventional formulation, obtaining full state information has a processing and measurement requirement that scales exponentially with the size of the Hilbert space~\cite{mohseni2008estimation, dasilva2011characterize}. However, physical states of interest typically have some structure that we can exploit to simplify the measurement complexity. Direct fidelity estimation (DFE) is a technique that utilizes this, and has been applied to matrix product states or stabilizer states in many-qubit systems
~\cite{cao2023qubits51,flammia2011directfidest,dasilva2011characterize,zhu2022estimate} 
to efficiently produce partial information about the system state. In the remainder of this work, we refer to such states as DFE-efficient.

Efficient QST is especially relevant in continuous variable systems with bosonic cavity modes, whose Hilbert spaces are arbitrarily large. These systems have applications in error correction codes~\cite{ofek2016extending,ni2023breakevencode,sivak2023breakevenqec}, quantum optics~\cite{sundaresan2015beyond}, quantum simulation~\cite{owens2018quarter}, and quantum information processing~\cite{naik2017random}. For a single mode, full state information is obtained by measuring operators like the Wigner operator~\cite{cahill1969wigner} or Q function operator at different mode displacements~\cite{shen2016optimizedtom}. Efficient QST in the multimode case is much more challenging.
Several efforts use multiple cavities to propose or produce increasingly complex states like multimode cat states~\cite{albert2019paircat}, W states~\cite{chakramhe2022blockade}, multimode GKP states~\cite{royer2022gridstates}, GHZ states~\cite{baumer2023ghz}, and other multimode Fock state superpositions~\cite{ouyang2019code,niu2018bosoniccode} that have a variety of applications in quantum error correction and logical encodings, as well as quantum simulation. In particular, W states have unique multipartite entanglement and protection against photon loss that gives them applications in quantum communication.
Some proposed theoretical methods are able to extract multimode state information while circumventing the exponential scaling of observation number with the number of modes. These include techniques that apply additional unitaries between modes as part of the measurement process, perform targeted measurements with polynomial post-processing~\cite{cramer2010efficienttom}, make use of ancillary modes and a known excitation number~\cite{chen2023trappediontom}, or apply operators based on excitation counting~\cite{shen2016optimizedtom}. To obtain more full state information, we instead directly measure the density matrices of potentially mixed states confined in a subspace of interest.

In this work, we use the Direct Extraction of (Density) Matrix Elements from Subspace Sampling Tomography (DEMESST) method to \textit{demystify} and reconstruct quantum state density matrices. 
DEMESST applies when an unknown state lies in a polynomial dimensional subspace. It only requires local operations if the subspace is spanned by a set of finite local operations acting on a DFE-efficient state~\cite{yatming}. Under these conditions, DEMESST has a polynomially scaling sampling requirement, and applies to both discrete qubit and continuous cavity systems.
For certain multimode cavity states, the total measurement number will therefore depend polynomially on the number of modes, rather than exponentially. This is especially advantageous when the subspace that an expected state lives in is much smaller than the full space. With DEMESST, we individually sample measurements for each basis operator in a polynomial subspace, and subsequently reconstruct a density matrix by combining them. Additionally, we implement this method with Wigner operators and Wigner tomography, thus performing the measurements with purely local operations on the modes and eliminating errors that may be associated with multimode unitaries or beamsplitters. With this approach, we measure fidelities of W states prepared in up to 4 bosonic modes, which is beyond existing demonstrations and advances the state of the art.

The structure of this paper is as follows: 
we first describe the details of DEMESST and elaborate on how it operates. We then discuss existing QST methods for continuous variable systems and how they have been optimized~\cite{philreinhold,chakramhe2022blockade}.
We compare DEMESST to these previous methods by testing the observation number required to accurately reconstruct the density matrices of approximate entangled multimode W states of 2, 3, and 4 modes on a superconducting cQED system, without making prior assumptions about the populations or phases of the state components. Finally, we experimentally confirm the advantage of DEMESST and verify its expected properties.
\begin{figure}[!t]
\includegraphics[width=0.483\textwidth]{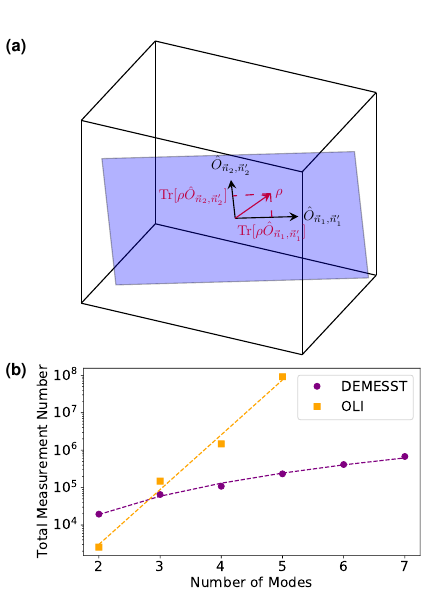}
\caption{\textbf{Theoretical comparison and schematic representation of tomography methods}. 
(a) Schematic representing the DEMESST method. Rather than sampling an entire multimode operator space (3D space), if a state lives in some number of subspaces (blue 2D plane), we restrict the sampling to each of those instead. The \{$O$\} basis operators are of the form $O_{\vec{n}, \vec{n}'} = \ket{\vec{n}}\bra{\vec{n}'}$ for generic basis states $\ket{\vec{n}}, \ket{\vec{n}'}$ (see supplementary information). Assuming an orthonormal basis, the state $\rho$ is given by $\rho = \sum_{\vec{n}_1, \vec{n}_2} \text{Tr}[\rho O_{\vec{n}_2, \vec{n}_1}] O_{\vec{n}_1, \vec{n}_2}$. This improves the overall efficiency of the sampling, especially for states with support across large numbers of modes. In practice, we use Hermitian $\{O\}$ that are accessible through experiment. 
(b) Number of measurements required for the DEMESST (purple, circles) and OLI (orange, squares) methods to reach a 90\% state reconstruction fidelity on W states of up to 7 modes, assuming perfect state preparation. OLI scales exponentially with the size of the Hilbert space (and therefore the number of modes $M$), while DEMESST scales only polynomially. 
} 
\label{Fig1}
\end{figure}

DEMESST scales polynomially with mode number for states that have support in a polynomial subspace of DFE-efficient basis operators~\cite{cao2023qubits51} (for example, states with a known maximum excitation number in the Fock basis).
This is accomplished by leveraging that information, and rather than sampling displacements corresponding to all basis operators, only sampling for the ones that are expected to support the state. This is illustrated schematically in Fig.~\ref{Fig1}(a). We reconstruct a density matrix $\rho$ in a polynomial subspace by independently measuring basis operators and the corresponding matrix elements for each projection of $\rho$ within the subspace, through methods similar to DFE~\cite{cao2023qubits51,flammia2011directfidest}, without introducing bias (see supplementary information). The reconstructed state is given by
\begin{equation}
    \rho = 
    \begin{bmatrix}
        \text{Tr}[\rho O_{\vec{n}_1, \vec{n_1}}] & \text{Tr}[\rho O_{\vec{n}_2, \vec{n}_1}] & \dots \\
        \hspace{4pt} & \text{Tr}[\rho O_{\vec{n}_2, \vec{n}_2}] \\
        \vdots & \hspace{4pt} & \ddots
    \end{bmatrix}
\end{equation}
With this approach, we avoid extracting irrelevant information about states outside the subspace of interest, thereby lowering the number of measurements required for an accurate result (see supplementary information). 
For example, if we know a 3-mode state has a maximum of 2 photons ($M = 3, N=3$ for 0, 1, or 2 photon population), we reconstruct it with DEMESST by measuring the matrix elements associated with that subspace, namely the one formed by $\ket{000}, \ket{001}, \ket{011}, \ket{002}$, and permutations. We eliminate unnecessary sampling of unnecessarily high photon number states like $\ket{111}$ or $\ket{012}$ and beyond. Additionally, we build upon existing work~\cite{cao2023qubits51,dasilva2011characterize} by developing and applying this approach to QST of bosonic modes and arbitrary mixed states rather than qubits and pure states.
Here, we implement DEMESST on a multimode cavity system with Wigner tomography.

\begin{figure*}
\includegraphics[width=0.96\textwidth]{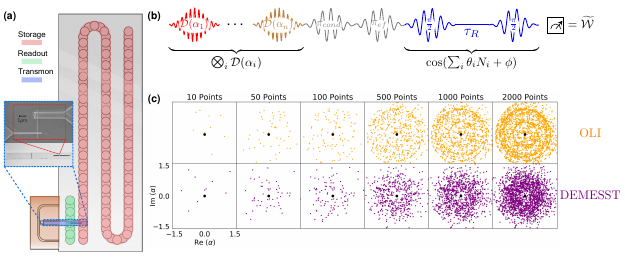}
\caption{\textbf{Experimental system and scheme}. (a) Diagram of the multimode flute cavity with a storage cavity and readout cavity that both couple to a transmon chip. Experimental drives are input through the readout cavity, or through a direct drive on the storage cavity. (b) Wigner tomography pulse sequence. Initial cavity displacements and a final generalized multimode parity measurement implement the tomography, while optional conditional pi pulses are used for projecting targeted modes out of the transmon ge subspace for the DEMESST approach (see supplementary information). An additional angle $\phi$ is applied between the $\pi/2$ pulses of the parity measurement to project it onto the real axis. (c) Cavity displacement plots for the OLI (orange) and DEMESST (purple) sampling methods. The OLI has ring features corresponding to measurement of all Fock states up to a cutoff, while DEMESST has points more densely located in the phase space based on the basis state being measured.} 
\label{Fig2}
\end{figure*}

Wigner tomography uses measurements of the Wigner operator $\mathcal{W}(\vec{\alpha}) = \mathcal{D}(\vec{\alpha}) \Pi \mathcal{D}(-\vec{\alpha})$ acting on a bosonic state $\rho$ to reconstruct it. Here, $\mathcal{D}(\vec{\alpha}) = \bigotimes_i \mathcal{D}(\alpha_i)$ is the displacement operator and $\Pi$ is a parity measurement. Existing inversion-based Wigner tomography methods operate by taking the Wigner functions of a set of displacements $\{\vec{\alpha}\}$ to construct a measurement matrix $\mathcal{M}$ that maps to states as $\vec{x} = \mathcal{M} |\rho \rangle \rangle$, where $|\rho \rangle \rangle$ is the vectorized form of $\rho$.
Inverting $\mathcal{M}$ to $ \minimize_{\rho}||\mathcal{M}|\rho \rangle \rangle - \vec{x}||$ allows us to determine the physical (unit trace and positive semidefinite) $\rho$ that was most likely to have produced $\vec{x}$. The set $\{\vec{\alpha}\}$ is optimized by minimizing the condition number (the ratio of largest to smallest eigenvalue) of $\mathcal{M}$ and thus the error magnification, using the techniques presented in~\cite{philreinhold,chakramhe2022blockade}. We refer to this method as Optimized Linear Inversion (OLI). In this approach, to make the problem tractable, we choose a cutoff dimension $N$ to truncate the Hilbert space. Reconstructing $\rho$ for a single mode therefore requires at least $N^2$ measurements to determine each density matrix parameter. 
For multiple modes, the size of the Hilbert space and thus the number of required measurements will scale exponentially, requiring at least $N^{2M}$ observations for $M$ modes. We compare OLI with DEMESST by testing their performance on experimentally prepared W states.

\bgroup
\def\arraystretch{1.5}
\begin{table*}
\centering
\begin{tabular}{|c|c|c|c|c|}
\hline
& Simulated Fidelity & OLI Fidelity & DEMESST Fidelity & OLI:DEMESST Distance \\
\hline
2-mode & 0.971 & 0.966(5) & 0.96(1) & 0.96(2) \\
\hline
3-mode & 0.956 & 0.949(4) & 0.955(4) & 1.55(10) \\
\hline
4-mode & 0.912 & 0.912(7) & 0.911(7) & 2.3(2) \\
\hline
\end{tabular}
\caption{\textbf{OLI and DEMESST Tomography results}. Simulations and Wigner tomography measurements are performed on $M$-mode W states of varying size. The fidelities are in good agreement, and the Frobenius norm matrix distance ratio demonstrates exponential improvement as $M$ increases. }
\label{results_table}
\end{table*}
\egroup

W states are excellent candidates for testing our Wigner tomography sampling methods.
For 2 modes, this forms a dual rail encoding, $\ket{\text{W}_2} = (\ket{10} + e^{i \phi} \ket{01}) / \sqrt{2}$. For 3 modes, $\ket{\text{W}_3} = (\ket{100} + e^{i \phi_1} \ket{010} + e^{i \phi_2} \ket{001}) / \sqrt{3}$, and similarly for 4 modes, $\ket{\text{W}_4} = (\ket{1000} + e^{i \phi_1} \ket{0100} + e^{i \phi_2} \ket{0010} + e^{i \phi_3} \ket{0001}) / \sqrt{4}$. Here, the $\phi_j$'s are \textit{a priori} unknown phases on each of the state components, and are determined through measurement. Additionally, due to imperfect state preparation, we make no assumptions about the component populations and precisely measure each one separately, making our measurement task more difficult. We further generalize this: rather than restricting ourselves to the pure states $\text{W}_j = \ket{\text{W}_j}\bra{\text{W}_j}$, we measure the full, possibly mixed density matrices. W states are great representative states because they are irreducible multimode states that generalize straightforwardly to arbitrary numbers of modes, and have a well-defined photon number. 
We prepare them easily using photon blockade~\cite{heeres2015cavity,heeres2017implementing,bretheau2015quantum,chakramhe2022blockade}.

We first investigate the simulated theoretical performance of DEMESST and OLI on $M$-mode W states. Assuming perfect state preparation, we compare the number of observations required to accurately reconstruct the W state with 90\% fidelity. This is shown in Fig.~\ref{Fig1}(b). 
As discussed previously, inversion-based methods like OLI have a sampling requirement that scales exponentially with mode number $M$.
In contrast, the DEMESST method scales polynomially with the subspace dimension and thus $M$, demonstrating an advantage that increases with $M$. For two modes, OLI performs better due to overhead required for the DEMESST approach (see supplementary information). However, for larger $M$, DEMESST requires fewer measurements to converge to the same level of fidelity, and scales much more efficiently than OLI. We proceed to demonstrating this expected behavior in experiment.

We generate W states and implement DEMESST and OLI on a superconducting 3D cQED platform. The system consists of a transmon qubit coupled to a 3D readout cavity and a 3D multimode storage cavity. A schematic of this hardware setup is shown in Fig.~\ref{Fig2}(a). The single storage cavity supports many bosonic cavity modes at roughly equally spaced microwave frequencies.
The transmon allows for universal control of the cavity modes and also mediates interactions like photon blockade~\cite{bretheau2015quantum,heeres2015cavity} between the storage modes. We use four of the modes to prepare our multimode W states. We also use the transmon to implement the parity measurements necessary for the multimode Wigner tomography.

The tomography sampling methods use the same generalized Wigner tomography sequence, where each cavity mode is displaced before performing a generalized parity measurement on the transmon~\cite{chakramhe2022blockade}. This procedure allows us to perform multimode tomography measurements despite having unequal dispersive shifts $\chi_m$ for different modes, without requiring $\chi$ engineering techniques~\cite{wang2016schrodinger,rosenblum2018fault} or additional control pulses. This pulse sequence is shown in Fig.~\ref{Fig2}(b).
For the DEMESST method, we may also include a $\pi_{ge}$ pulse conditioned on certain cavity populations followed by an $\pi_{ef}$ pulse on the transmon. 
These pulses project one or more of the cavity modes to the transmon $\ket{f}$ level and outside the $\ket{g}-\ket{e}$ qubit space of the Wigner tomography. This effectively removes those modes from the measurement and allows us to perform it more easily. We apply this when the basis operator being measured has one or more cavity modes in vacuum. For example, to sample the 3-mode matrix element $\ket{001}\bra{010}$, we project the first mode to $\ket{f}$ and reduce the sampling to the 2-mode $\ket{01}\bra{10}$. This projection allows us to reduce the size of the sampling problem for that element to a lower number of modes (see supplementary information).

\begin{figure*}
\includegraphics[width=0.99\textwidth]{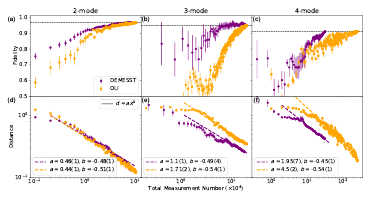}
\caption{\textbf{Tomography fidelity and matrix distances for DEMESST and OLI sampling methods}. Sampling was performed on approximate entangled W states of 2--4 modes. The top panels (a)--(c) show fidelities versus~an ideal (exactly equal population coefficients) W state for 2--4 modes, with dashed horizontal lines indicating the final converged fidelity obtained from the OLI method. These final fidelities are, for 2--4 modes, $0.966 \pm 0.005, 0.949 \pm 0.004$, and $0.912\pm0.007$ for OLI and $0.96\pm0.01, 0.954\pm0.004$, and $0.911\pm0.007$ for DEMESST and are in good agreement. The bottom panels (d)--(f) show Frobenius norm matrix distances between the state at a given measurement number versus the final measured state. The rates of convergence are close to 1/$\sqrt{x}$ or a power of -0.5, as expected. As the mode number increases, the DEMESST method performs increasingly more efficiently by requiring fewer measurements to reach a given level of convergence or error threshold.
} 
\label{Fig3}
\end{figure*}

We first compare the state reconstructed from the OLI method with simulation, which we use as a baseline for later comparison to DEMESST. With OLI, we find prepared W state fidelities that are in good agreement with the simulated fidelities, as shown in Table~\ref{results_table}.
The simulations include error from transmon and cavity decoherence and decay, and state preparation errors such as leakage outside of the blockaded subspace. We now continue to the DEMESST performance.

For the DEMESST approach, we reconstruct the 2--4 mode density matrices by measuring Wigner operators corresponding to multimode Fock basis states with up to 2 photons. 
Even though W states have at most one photon, we measure two-photon operators to capture imperfect state preparation errors. These observations directly provide the values of each density matrix element. From the density matrix, we obtain the component populations and phase angles $\phi_j$ of our prepared approximate W states by calculating the phase angle value that best matches the resulting data. These angles are then verified to match with the ones obtained from the OLI approach.

We quantify the performance of the DEMESST and OLI sampling methods versus total measurement number with two metrics: reconstructed state fidelity and Frobenius norm matrix distance. The fidelities are computed with respect to an ideal W state, while the matrix distances are calculated with respect to the experimentally prepared state reconstructed at the maximum measurement number.
These results are shown in Fig.~\ref{Fig3}. The final fidelities obtained from the DEMESST approach are presented in Table~\ref{results_table}, and
are consistent with the OLI results. Error bars are obtained from the results of 10 independent sets of 10 repetitions of tomography measurements for each sampled displacement. The number of distinct displacements for the OLI method therefore equals the Total Measurement Number shown on the x-axis in Fig.~\ref{Fig3} divided by 10. For the DEMESST method, the number of distinct displacements for each basis element is further divided by the number of elements. For the 2-mode W state, the two methods perform similarly, while for 3 and 4 modes, DEMESST performs better than OLI with faster convergence to the final state fidelity. 

This improvement is most evident in the matrix distance comparisons. The distances are computed using the Frobenius norm, with error bars again computed from 10 independent sets of 10 repetitions for each displacement. The final state density matrix against which the distances are computed is obtained by considering all 100 repetitions, rather than sets of 10, which is why the final distances do not completely vanish.
The behavior of both sampling methods is nearly identical for the 2-mode W state. However, for the 3-mode case, the DEMESST has noticeably faster convergence versus total measurement number $x$, as the matrix distance $d$ to the final state is smaller, as seen by the fit coefficient to $d = ax^b$ ($a=1.1\pm0.1$ for DEMESST versus~$1.71\pm0.02$ for OLI). This effect is further enhanced in the 4-mode case. The ratio of these fit values scales roughly geometrically, as shown in Table~\ref{results_table}, 
reflecting the fact that OLI scales exponentially while DEMESST only scales polynomially versus total measurement number. In all cases, the distances fall off roughly as $x^{-1/2}$, as expected.

An advantage of the DEMESST sampling method compared to OLI is its self-consistency. Individual density matrix elements for any multimode state are measured independently, without needing to choose a cutoff maximum photon number or Hilbert space size that could subject the reconstructed state to inversion errors. This eliminates the risk of obtaining an inaccurate tomography result if, for example, the prepared state contains population beyond the space spanned by our chosen basis during OLI sampling.

\begin{figure*}
\includegraphics[width=0.99\textwidth]{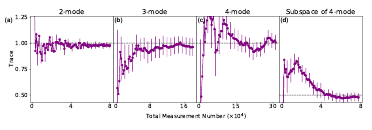}
\caption{
\textbf{Trace verification for the DEMESST sampling method}. Trace versus point number for prepared (a) 2-mode, (b) 3-mode, and (c) 4-mode W states. Error bars are shown for every fourth point, as well as the final one. They are obtained from comparing the traces from individual sets of measurements. As expected, the traces converge to values near unity. (d) Trace versus point number when measuring only a 2-mode subspace of a prepared 4-mode W state. Due to only measuring half of the populated space, the trace converges to 0.5. This demonstrates that the DEMESST sampling method is self-consistent and does not depend on the chosen measurement subspace.
} 
\label{Fig4}
\end{figure*}

We verify that the DEMESST tomography sampling method leads to self-consistent measurement results. We check the traces of our prepared W states and compare them with unity, the expected value for physical states. This allows us to confirm that our prepared state indeed lives in our chosen, measured Hilbert space.
The results are shown in Fig.~\ref{Fig4} as average trace versus observation number. Like before, the averages are taken over 10 independent sets of 10 measurement repetitions for each sampled displacement. We find that in all cases ((a)--(c)), the observed traces are near one.
We attribute large deviations from unity at low measurement numbers to noise and statistics, and attribute the final traces being slightly less than one to imperfect state preparation that produces population outside the measured subspace.
We perform a further check by considering only a 2-mode subspace of a prepared 4-mode W state. This is shown in Fig.~\ref{Fig4}(d). As expected, the measured trace converges to a value near 0.5, as we are effectively only observing half of the total state population. This demonstrates that the DEMESST method provides accurate results for each basis operator independently. In particular, we can identify when we have measured insufficient basis elements to fully characterize a state, such as when the state lives partially (or entirely) outside the corresponding space, which is a useful capability in itself. 

In summary, we have applied the DEMESST sampling method to characterize multimode cavity states with Wigner tomography. DEMESST is most appropriate for multimode states that have population contained in a small subspace of DFE-efficient elements of an overall Hilbert space, and outperforms traditional optimized inversion-based methods by scaling polynomially rather than exponentially with mode number. We observe this improvement for W states on 3 and 4 modes. Here, we have presented comparisons using the multimode Fock basis on multimode W states, but DEMESST also applies to different bases that more readily support other states; this tomography method can even be used for DFE by choosing as a basis the intended target state. While Wigner tomography was presented in this work, the method also operates beyond the bosonic Wigner function, and works for both continuous and discrete systems. This approach operates without coupling gates between modes, and would be useful for calibrating entangled states over distributed quantum networks.
Ultimately, the DEMESST sampling method enables efficient measurement of large states, which will be crucial as the size of quantum hardware systems increases and more complicated states are generated and applied for quantum simulation, bosonic logical state encoding, and error correction.

\begin{acknowledgments}
K.H., D.I.S. acknowledge support from the Samsung Advanced Institute of Technology Global Research Partnership. This work was also supported by ARO Grants W911NF-15-1-0397 and W911NF-16-1-0349, AFOSR MURI grant  FA9550-19-1-0399, and the Packard Foundation (2013-39273). This work is funded in part by EPiQC, an NSF Expedition in Computing, under grant CCF-1730449. D.I.S. acknowledges support from the David and Lucile Packard Foundation. This work was partially supported by the University of Chicago Materials Research Science and Engineering Center, which is funded by the National Science Foundation under award number DMR-1420709. Devices were fabricated in the Pritzker Nanofabrication Facility at the University of Chicago, which receives support from Soft and Hybrid Nanotechnology Experimental (SHyNE) Resource (NSF ECCS-1542205), a node of the National Science Foundation’s National Nanotechnology Coordinated Infrastructure.
M.Y., Y.W., L.J. acknowledge support from the ARO (W911NF-23-1-0077), ARO MURI (W911NF-21-1-0325), AFOSR MURI (FA9550-19-1-0399, FA9550-21-1-0209), AFRL (FA8649-21-P-0781), NSF (OMA-1936118, ERC-1941583, OMA-2137642), NTT Research, and the Packard Foundation (2020-71479). A.S. acknowledges support by a Chicago Prize Postdoctoral Fellowship in Theoretical Quantum Science. L.J. acknowledges the support from the Marshall and Arlene Bennett Family Research Program. This material is based upon work supported by the U.S. Department of Energy, Office of Science, National Quantum Information Science Research Centers. 
\end{acknowledgments}

% \putbib

% \begin{center}
%     \textbf{\small REFERENCES}
% \end{center}
% \printbibliography[heading=none]
% \end{refsection}
% \end{bibunit}

\clearpage
%%%%%%%%%%%%%%%%%%%%%%%%%%%%%%%%%%%%%%%%%%%%%%%%%%%%%%%%%%
% \appendix

\definecolor{Gray}{gray}{1}
\definecolor{storagecolor}{rgb}{1,0.97,0.9}
\definecolor{tmoncolor}{rgb}{0.92,1,1}
\definecolor{readoutcolor}{rgb}{1,0.9,0.9}
\setcounter{secnumdepth}{1}

\newcommand{\Tr}{\text{Tr}}
\newcommand{\dd}{\text{d}}
\newcommand{\spl}{\text{spl}}

\setcounter{equation}{0}
\setcounter{figure}{0}
\renewcommand{\thefigure}{S\arabic{figure}}

\begin{widetext}
\begin{center}
\textbf{\large Supplementary Information: Efficient multimode Wigner tomography}
\end{center}
\end{widetext}
% \begin{refsection}
% \bibliographyunit[\chapter]
% \begin{bibunit}

\section{Cryogenic setup and control instrumentation}
\begin{figure}[h]
  \begin{center}
    \includegraphics[width=0.48\textwidth]{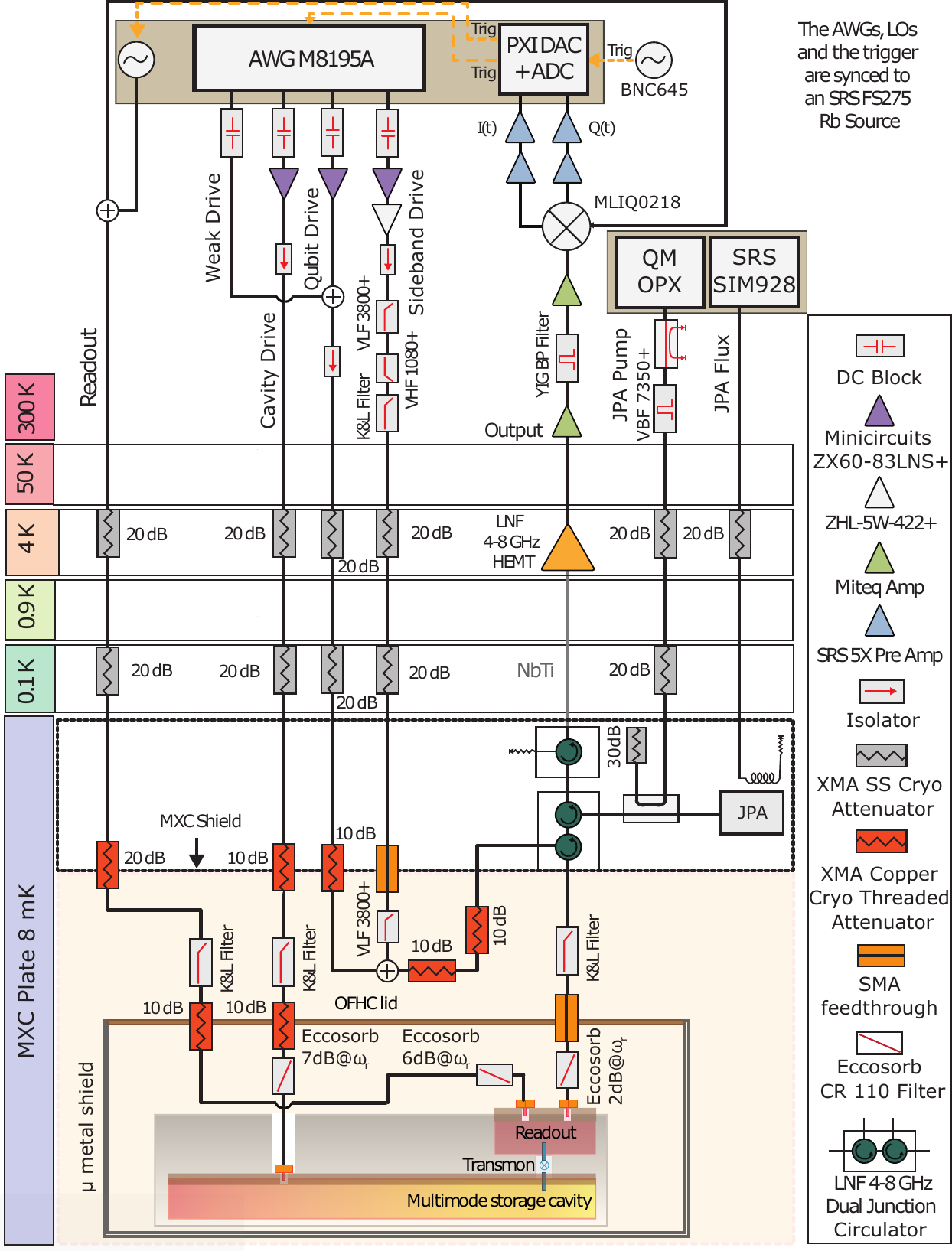}
    \caption{Schematic of the cryogenic setup, microwave wiring and filtering, and control instrumentation.}
  	\label{wiring diagram}
  \end{center}
\end{figure} 
The multimode cavity device is heat sunk to an OFHC copper plate connected to the base stage of a Bluefors LD-400 dilution refrigerator (10-12 mK). The sample is surrounded by a can containing two layers of $\mu$-metal shielding, with the inside of the inner layer connected to a can made out of copper shim that is painted on the inside with Berkeley black and attached to the copper can lid. A schematic of the cryogenic setup, control instrumentation, and device wiring is shown in SFig.~\ref{wiring diagram}. The device is machined from a single piece of 5N5 aluminium and consists of a readout cavity and a multimode storage cavity fabricated using the flute method described in~\cite[S][]{chakramoriani2020} and also used in~\cite[S][]{chakramhe2022blockade2}. The cavities are bridged by a 3D transmon circuit. Controls are performed through the readout cavity or the direct cavity drive line, by driving at the qubit and storage mode frequencies. The pulses are directly digitally synthesized using a four-channel, 64 GSa/s arbitrary waveform generator (Keysight M8195A). The combined signals are sent to the device after being attenuated at each of the thermal stages, as shown in SFig.~\ref{wiring diagram}. 
Outside the fridge, the signal is filtered (tunable narrow band YIG filter with a bandwidth of 80 MHz) and further amplified. The amplitude and phase of the resonator transmission signal are obtained through a homodyne measurement, with the transmitted signal demodulated using an IQ mixer and a local oscillator at the readout resonator frequency. The homodyne signal is amplified (SRS preamplifier) and recorded using a fast ADC card (Keysight M3102A PXIe 500 MSa/s digitizer).

\section{Calibration of the multimode Hamiltonian}
The Hamiltonian of the multimode cQED system realized by the transmon and the storage modes is:
\begin{equation}
\begin{split}
H &=   \omega_q \ket{e}\bra{e} + \sum_{m=0}^{N-1}\{\omega_m a_m^{\dagger} a_m + \chi_m a_m^{\dagger} a_m \ket{e}\bra{e} \, \\ 
&+ \frac{k_m}{2} a_m^{\dagger} a_m (a_m^{\dagger} a_m - 1)\} + \sum_{n \neq m} k_{mn} a_m^{\dagger} a_m a_n^{\dagger}a_n ,
\end{split}
\label{seqn1}
\end{equation}
where $\omega_q/(2\pi) = 4.96$ GHz is the frequency of the transmon $\ket{g} \mhyphen \ket{e}$ transition, $\omega_{m}/(2\pi)$ ranging from $5.72 - 6.47$ GHz are the cavity mode frequencies, $\chi_m/(2\pi)$ ranging from $-1.64$ to $-0.91$ MHz are the dispersive shifts, $k_m$ the self-Kerr shift of each mode, and $k_{mn}$ the cross-Kerr interactions between the modes. The Kerr nonlinearities range from $-6 - 7$ kHz. Parameter values are determined with the processes described in the supplement of~\cite[S][]{chakramhe2022blockade2}. A summary of the system parameters used in the experiment, as well as Liouvillian terms corresponding to transmon and cavity decoherence and decay, is provided in STable~\ref{parameters_table}. The 2-mode experiments used the last two modes in the table, the 3-mode experiments used the last three modes, and the 4-mode experiments used all four of the modes.

\begin{table*}
% \begin{center}
\setlength\arrayrulewidth{1.3pt}
\def\arraystretch{1.3}  % change spacing, default is 1
\begin{tabular}{|c|c|c|c|}
\hline
\rowcolor{Gray}
Parameter & Hamiltonian/Liouvillian Term & Quantity & Value(s) \\
\hline
\rowcolor{tmoncolor}
Transmon frequency & $\omega_q \ket{e}\bra{e}$ & $\omega_q/(2\pi)$ & 4.97 GHz \\
\hline
\rowcolor{storagecolor}
Storage cavity frequencies & $\omega_m a_m^{\dagger} a_m$ & $\omega_m/(2\pi)$ & 5.716, 5.965, 6.223, 6.469 GHz \\
\hline
\rowcolor{readoutcolor}
Readout frequency &  $\omega_r a_r^{\dagger} a_r$ & $\omega_r/(2\pi)$ & 7.79 GHz \\
\hline
\rowcolor{readoutcolor}
Readout dispersive shift & $\chi_r a_r^{\dagger} a_r \ket{e}\bra{e}$ & $\chi_r/(2\pi)$ & 1 MHz \\
\hline
\rowcolor{storagecolor}
Storage mode dispersive shifts & $\chi_m a_m^{\dagger} a_m \ket{e}\bra{e}$ & $\chi_m/(2\pi)$ & -1.636, -1.269, -1.093, -0.906 MHz \\
\hline
\rowcolor{storagecolor}
Storage mode self-Kerrs & $\frac{k_m}{2} a_m^{\dagger} a_m (a_m^{\dagger} a_m - 1)$ & $k_m/(2\pi)$ & 9.0, 5.2, 4.2, 3.2 kHz \\
\hline
\rowcolor{storagecolor}
Storage mode cross-Kerrs & $k_{mn}a_m^{\dagger} a_m a_n^{\dagger} a_n $ & $k_{mn}/(2\pi)$ & $-6 - 0$ kHz \\
\hline
\rowcolor{tmoncolor}
Transmon $\ket{e} \rightarrow \ket{g}$ relaxation & $\frac{1}{T^q_1}(1+\bar{n}) \mathcal{D}\big[\ket{g}\bra{e}\big]$ & $T^q_1$ & $ 108 \pm 7~\mu$s  \\
\hline
\rowcolor{tmoncolor}
Transmon $\ket{g}-\ket{e}$ dephasing & $\big(\frac{1}{T^q_2}-\frac{1}{2T^{q}_{1}}\big) \mathcal{D}\big[\ket{e}\bra{e}\big]$ & $T^q_2$ & $ 165 \pm 14~\mu$s \\
\hline
\rowcolor{readoutcolor}
Readout linewidth &  $\kappa_r \mathcal{D}[a_r]$ & $\kappa_r/(2\pi)$ & 0.52 MHz \\
\hline
\rowcolor{storagecolor}
Storage mode relaxation & $\frac{1}{T^m_1} \mathcal{D}[a]$ & $T^m_1$ & $\sim 1-2$ ms, see~\cite[S][]{chakramoriani2020} \\
\hline
\rowcolor{tmoncolor}
Transmon thermal population & $\frac{\bar{n}}{T^q_1} \mathcal{D}\big[\ket{e}\bra{g}\big]$ & $\bar{n}$ & $1.5 \pm 0.5$ \% \\
\hline
\rowcolor{storagecolor}
Storage mode dephasing & $\big(\frac{1}{T^m_2}-\frac{1}{2T^{m}_{1}}\big) \mathcal{D}\big[\ket{1}\bra{1}\big]$ &  $T^m_2$ &$\sim 1.5-3$ ms, see~\cite[S][]{chakramoriani2020}\\
\hline

\end{tabular}
\captionsetup{justification=centering}
\caption{\textbf{Multimode cQED system parameters}}
\label{parameters_table}
% \end{center}
\end{table*}

\section{The Wigner function and its generalization}

The Wigner function is the quasiprobability distribution of a state in phase space, and is one of the most important functions in the field of quantum optics. For a single-mode state $\rho$, the Wigner function is defined as~\cite[S][]{cahill_density_1969}
\begin{equation} \label{eq: mode-1-wig}
    W_\rho(\alpha) = 2\Tr[\rho D(\alpha) e^{i\pi a^\dag a}D^\dag(\alpha)],
\end{equation}
where $D(\alpha) = e^{\alpha a^\dag - \alpha^* a}$ is the displacement operator. We can see that $W_\rho(\alpha)$ is proportional to the expectation value of the parity operator with the state $\rho$ displaced by complex amplitude $-\alpha$. Similarly, we can introduce the Wigner function for an operator $O$ with finite Frobenius norm (F-norm) ($||O||_F = \sqrt{\Tr[O^\dag O]} < +\infty$) by substituting the state $\rho$ with the operator $O$ in Eqn.~\eqref{eq: mode-1-wig}.

The Wigner function of a multimode $M$-mode state $\rho$ can be expressed as
\begin{equation} \label{eq: mode-M-wig}
    W_\rho(\vec{\alpha}) = 2^M \Tr[\rho D(\vec{\alpha}) e^{i\pi \sum_{m=1}^M a_m^\dag a_m}D^\dag(\vec{\alpha})],
\end{equation}
where $D(\vec{\alpha}) = \prod_{m=1}^M e^{\alpha_m a_m^\dag - \alpha_m^* a_m}$. As described in the main text, due to the different dispersive shifts of our cavity modes, we instead implement a generalized parity operator~\cite[S][]{chakramhe2022blockade2}. We can express the corresponding
generalized version of the Wigner function, as introduced in~\cite[S][]{cahill_ordered_1969}, as
\begin{equation} \label{eq: mode-M-genwig}
    W_\rho(\vec{\alpha}, \vec{\theta}) = \frac{2^M \Tr[\rho D(\vec{\alpha}) e^{i \sum_{m=1}^M \theta_m a_m^\dag a_m}D^\dag(\vec{\alpha})]}{\prod_{m=1}^M [1+i\cot(\theta_m/2)]},
\end{equation}
where $\theta_m$ can be different and need not equal $\pi$ for the $M$ modes. We also denote
\begin{equation}\label{eq: tilde-genwig}
\tilde{W}_\rho(\vec{\alpha}, \vec{\theta}) = \Tr[\rho D(\vec{\alpha}) e^{i \sum_{m=1}^M \theta_m a_m^\dag a_m}D^\dag(\vec{\alpha})],
\end{equation}
where $\tilde{W}_\rho(\vec{\alpha}, \vec{\theta})$ can in general be a complex number, since $e^{i \sum_{m=1}^M \theta_m a_m^\dag a_m}$ is no longer Hermitian. Also, $|\tilde{W}_\rho(\vec{\alpha}, \vec{\theta})| \leq 1$. We can see from the form of Eqn.~\eqref{eq: tilde-genwig} that $\tilde{W}_\rho(\vec{\alpha}, \vec{\theta}) = \tilde{W}^*_{\rho^\dag}(\vec{\alpha}, -\vec{\theta})$. This generalized Wigner function $\tilde{W}_\rho(\vec{\alpha}, \vec{\theta})$ is what we experimentally measure. In the next section, we show that $\tilde{W}_\rho(\vec{\alpha}, \vec{\theta})$ for quantum states plays a similar role as the usual Wigner function (Eqn.~\eqref{eq: mode-M-wig}) when calculating the expectation values with other operators.

\section{Estimating expectation values with Wigner sampling}\label{sec: wig-sampling}

In this section, we discuss the sampling method for estimating the expectation value of the Wigner function with an operator with finite F-norm $O$. We then analyze the overhead required to reach a certain accuracy threshold. 

We start with the identity that reflects the relationship between expectation values and the Wigner function:
\begin{equation}
    \Tr[\rho O] = \int \frac{\dd^{2M} \vec{\alpha}}{\pi^M} \ W_\rho(\vec\alpha) W_O(\vec{\alpha}).
\end{equation}
For the generalized Wigner function, we have a similar expression:
\begin{equation}\label{eq: expct-genwig}
\begin{split}
    \Tr[\rho O] &= \int \frac{\dd^{2M} \vec{\alpha}}{\pi^M} \ W_\rho(\vec\alpha, -\vec\theta) W_O(\vec{\alpha}, \vec\theta)\\
    &= C_M\int \dd^{2M} \vec\alpha \ \tilde W_\rho(\vec\alpha, -\vec\theta) \tilde W_O(\vec{\alpha}, \vec\theta),
\end{split}
\end{equation}
where $C_M = \prod_{m=1}^M [2(1-\cos\theta_m)/\pi]$.

The equations above have been applied for direct fidelity estimation between an experimentally prepared state $\rho$ and a target pure state $\sigma$~\cite[S][]{da_silva_practical_2011}. For example, we can rewrite Eqn.~\eqref{eq: expct-genwig} as
\begin{equation}\label{eq: W2-method}
\begin{split}
    F(\rho, \sigma) &= \Tr[\rho \sigma] = C_M\int \dd^{2M} \vec\alpha \ |\tilde W_\sigma(\vec{\alpha}, \vec\theta)|^2 \frac{\tilde W_\rho(\vec\alpha, -\vec\theta)}{\tilde W_\sigma(\vec\alpha, -\vec\theta)}\\
    &= \int \dd^{2M} \vec\alpha \ p_{W^2}(\vec\alpha) \frac{\re[e^{i\phi(\vec\alpha)}\tilde W_\rho(\vec\alpha, -\vec\theta)]}{|\tilde W_\sigma(\vec\alpha, -\vec\theta)|},
\end{split}
\end{equation}
where $p_{W^2}(\vec\alpha) = C_M |\tilde W_\sigma(\vec{\alpha}, \vec\theta)|^2$ satisfies $p_{W^2}(\vec\alpha) \geq 0$ and $\int \dd^{2M} \vec\alpha \ p_{W^2}(\vec\alpha) = \Tr[\sigma^2] = 1$, such that we can treat $p_{W^2}(\vec\alpha)$ as a probability distribution function that we can then use to sample a set of displacement vectors $\{\vec\alpha^{(k)}\}$. 
Given a set $\{\vec\alpha^{(k)}\}$, we can measure $\re[e^{i\phi(\vec\alpha^{(k)})}\tilde W_\rho(\vec\alpha^{(k)}, -\vec\theta)]$ and calculate the average $\frac{\re[e^{i\phi(\vec\alpha^{(k)})}\tilde W_\rho(\vec\alpha^{(k)}, -\vec\theta)]}{|\tilde W_\sigma(\vec\alpha^{(k)}, -\vec\theta)|}$ to obtain an estimate of $F(\rho, \sigma)$. 
In Sec.~\ref{sec: exp-prtcl-anal}, we show explicitly that by repeatedly measuring whether the qubit is in the $\ket{g}$ level, we can obtain a series of binomial outcomes $A^{(k)}_j \in \{1, -1\}$ whose expectation values lead to exactly $\re[e^{i\phi(\vec\alpha^{(k)})}\tilde W_\rho(\vec\alpha^{(k)}, -\vec\theta)]$. 
We call the above sampling method the $W^2$ tomography sampling method.

In general, we have the freedom to choose other probability distribution functions $p(\vec\alpha)$ to generate sampling points $\{\vec\alpha^{(k)}\}$, since
\begin{equation}
\begin{split}
    &\Tr[\rho O]\\
    ={}& C_M\int \dd^{2M} \vec\alpha \ p(\vec\alpha)\frac{|\tilde W_O(\vec{\alpha}, \vec\theta)|}{p(\vec\alpha)}\re[e^{i\phi(\vec\alpha)}\tilde W_\rho(\vec\alpha, -\vec\theta)].
\end{split}
\end{equation}
We can calculate the average of $C_M\frac{|\tilde W_O(\vec\alpha^{(k)}, \vec\theta)|}{p(\vec\alpha^{(k)})}A^{(k)}_j$ to obtain an estimate for $\Tr[\rho O]$. However, in certain situations, there will be an optimal choice for $p(\vec\alpha)$. For example, if we perform single shot measurements where we only measure each sampling vector $\vec\alpha^{(k)}$ once, the variance of the estimator $C_M\frac{|\tilde W_O(\vec\alpha^{(k)}, \vec\theta)|}{p(\vec\alpha^{(k)})}A^{(k)}$ will be limited by
\begin{equation} \label{eq: 1-shot-var}
\begin{split}
    & C^2_M \int \dd^{2M} \vec\alpha \ p(\vec\alpha) \frac{|\tilde W_O(\vec{\alpha}, \vec\theta)|^2}{p^2(\vec\alpha)} - (\Tr[\rho O])^2\\
    \geq{}& \frac{[C_M \int \dd^{2M} \vec\alpha \ |\tilde W_O(\vec{\alpha}, \vec\theta)|]^2}{\int \dd^{2M} \vec\alpha \ p(\vec\alpha)} - (\Tr[\rho O])^2\\
    ={}& \left[C_M \int \dd^{2M} \vec\alpha \ |\tilde W_O(\vec{\alpha}, \vec\theta)|\right]^2 - (\Tr[\rho O])^2.
\end{split}
\end{equation}
Here, we have used the Cauchy-Schwarz inequality and the fact that $\int \dd^{2M} \vec\alpha \ p(\vec\alpha) = 1$. The minimum is achieved when $p(\vec\alpha) \propto |\tilde W_O(\vec{\alpha}, \vec\theta)|$, which we call the DEMESST sampling method. It is also worth noting that, for the $W^2$ method where $p_{W^2}(\vec\alpha) = C_M |\tilde W_O(\vec{\alpha}, \vec\theta)|^2$, the integral shown in the first line of Eqn.~\eqref{eq: 1-shot-var} will be divergent. To avoid this, Ref.~\cite[S][]{da_silva_practical_2011,flammia_direct_2011} have proposed choosing a threshold cutoff value to discard some sampling vectors $\vec\alpha^{(k)}$ that make the denominator of $\frac{A^{(k)}_j}{|\tilde W_O(\vec\alpha^{(k)}, \vec\theta)|}$ too small. This cutoff procedure can lead to bias when estimating $\Tr[\rho O]$ and makes the error analysis more complicated. A detailed analysis of the effect of this cutoff in a multimode setting is beyond the scope of this work.

Instead, we focus on the DEMESST method. The probability distribution function is given by 
\begin{equation}
    p_{\text{D}}(\vec\alpha) = \frac{|\tilde W_O(\vec{\alpha}, \vec\theta)|}{Z_O},
\end{equation}
where 
\begin{equation}
    Z_O = \int \dd^{2M} \vec\alpha \ |\tilde W_O(\vec{\alpha}, \vec\theta)|.
\end{equation}
In the DEMESST method, we must average $C_M Z_O A^{(k)}_j$ over all sampling vectors $\vec\alpha^{(k)}$ and all possible binomial outcomes from the qubit measurement per sampling vector. In the limit where we do one qubit measurement per $\vec\alpha^{(k)}$, we can use Hoeffding's inequality to estimate the number of samples $N_\spl$ required to reach an accuracy $\epsilon_1$ with probability $1 - \delta_1$ to be
\begin{equation}
    P\left(\left|\frac{C_M Z_O}{N_\spl} \sum_{k=1}^{N_\spl} A^{(k)} - \Tr[\rho O]\right| \geq \epsilon_1\right) \leq \delta_1
\end{equation}
when
\begin{equation} \label{eq: spl-oh}
    N_\spl \geq \lceil \frac{2C_M^2 Z_O^2}{\epsilon_1^2}\ln(2/\delta_1) \rceil.
\end{equation}
We can see that, in general, $N_\spl \propto (C_M Z_O)^2$. In the next section, we analyze the properties of $Z_O$ for our operators of interest.

\section{Density matrix reconstruction procedure}\label{sec: dm-recstr}

\begin{figure*}
  % \begin{center}
    \includegraphics[width=0.7\textwidth]{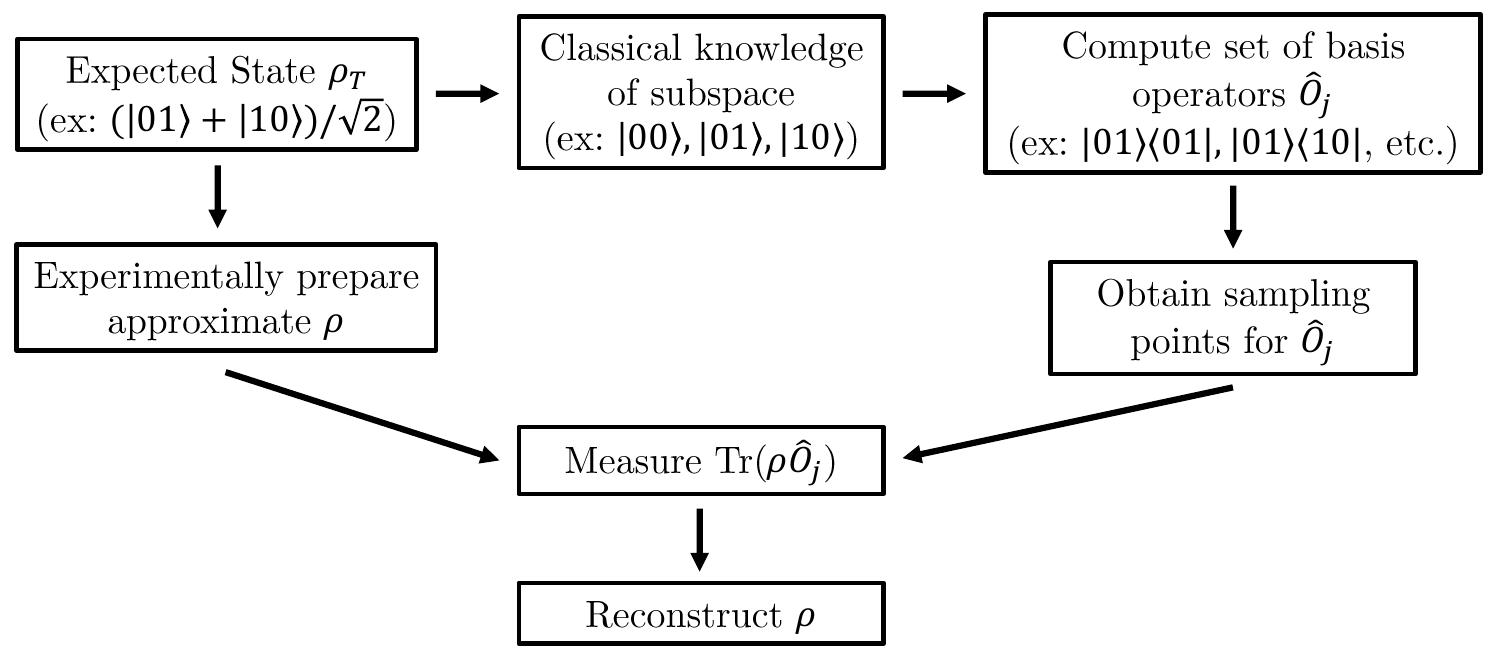}
    \caption{Flowchart depicting the steps involved in our density matrix reconstruction procedure, of which DEMESST is an example. Starting with a target expected state $\rho_T$, we can use our knowledge of $\rho_T$ to identify a target subspace and compute the operators $\hat{O}_j$ that span that basis. We then use those operators to generate sampling points, which may involve projections to be done efficiently, as described in Sec.~\ref{sec: wig-sampling}. By experimentally implementing the operators on our prepared state, we can reconstruct the state $\rho$ and compare it with the ideal target $\rho_T$. Even if the prepared state $\rho$ is not close to $\rho_T$, as long as it lives in the same subspace, we can effectively reconstruct it.}
  	\label{method_flowchart}
  % \end{center}
\end{figure*}

In this section, we discuss our scheme for reconstructing an unknown state using the DEMESST sampling method to estimate each element of its density operator in the Fock basis. We consider a system with $M$ modes and maximum total photon number $N$ between those modes. This restricts the dimension of the Hilbert space to $\binom{M+N}{N}$. We focus on the scaling of the sampling overhead (number of samples required) vs. mode number $M$, in the limit where $M$ is much larger than $2N$, and show that this overhead scales polynomially vs. $M$ with bounded photon number $N$, demonstrating the efficiency of the DEMESST approach as $M$ increases.

We assume that the operator $O$ (in $\Tr[\rho O]$) is in one of the following forms:
\begin{equation}\label{eq: ops-basis}
\begin{split}
    O_{\vec n, \vec n} &= \ketbra{\vec n}{\vec n},\\
    O^R_{\vec n, \vec n'} &= \frac{\ketbra{\vec n}{\vec n'} + \ketbra{\vec n'}{\vec n}}{\sqrt{2}} \quad (\vec n \neq \vec n'),\\
    O^I_{\vec n, \vec n'} &= i\frac{\ketbra{\vec n}{\vec n'} - \ketbra{\vec n'}{\vec n}}{\sqrt{2}} \quad (\vec n \neq \vec n').
\end{split}
\end{equation}
Here $\ket{\vec n} = \bigotimes_{m=1}^M \ket{n_m}$ and $\ket{\vec n'} = \bigotimes_{m=1}^M \ket{n'_m}$ are Fock basis states that satisfy $\sum_{m=1}^M n_m \leq N$ and $\sum_{m=1}^M n'_m \leq N$. The operators $O$ are chosen such that $O = O^\dag$ and $\Tr[O^\dag O] = 1$. For a system with $M$ modes and maximum total photon number $N$, we have $\binom{M+N}{N}^2$ of these operators.

One essential observation is that, when $M > 2N$, for any $(\vec n, \vec n')$ pair there are at least $(M-2N)$ elements in the set $S \equiv S_{(\vec n, \vec n')} = \{m|n_m = n'_m = 0\}$. We also denote $\bar S = \{1, 2, \dots, M\} \backslash S$. Because of this, the corresponding operators shown in Eqn.~\eqref{eq: ops-basis} can always be decomposed as
\begin{equation}\label{eq: op-decomps}
    O = \Big(\bigotimes_{m \in S} \ketbra{0}{0}_m\Big) \otimes O_{\bar S},
\end{equation}
where $O_{\bar S}$ has support on the modes with index $m\in \bar S$. We can see that the number of elements in $\bar S$ is no greater than $2N$, and independent of $M$. Similarly, the generalized Wigner function of such an operator $O$ satisfies
\begin{equation}
\begin{split}
    \tilde W_O(\vec{\alpha}, \vec\theta) ={}& \Big(\prod_{m \in S} \tilde W_{\ketbra{0}{0}}(\alpha_m, \theta_m)\Big) \\
    &\cdot \tilde W_{O_{\bar S}}(\vec\alpha_{\bar S}, \vec\theta_{\bar S}),
\end{split}
\end{equation}
where $\vec\alpha_{\bar S}$, $\vec\theta_{\bar S}$ contain those elements in $\vec\alpha$, $\vec\theta$ whose mode index $m \in \bar S$. Again, $\tilde W_{O_{\bar S}}(\vec\alpha_{\bar S}, \vec\theta_{\bar S})$ is independent of $M$.

Now, we consider the sampling overhead to obtain a precise estimate of $\Tr[\rho O]$ when $O$ satisfies the properties described above. One approach is to sample directly according to the $M$-mode function $|\tilde W_O(\vec{\alpha}, \vec\theta)|/Z_O$. Based on Eqn.~\eqref{eq: spl-oh}, the key quantity we should focus on is $C_M Z_O$, which satisfies
\begin{equation}
\begin{split}
    C_M Z_O ={}& \Big(\prod_{m=1}^M \frac{2(1-\cos\theta_m)}{\pi}\Big)\int \dd^{2M} \vec\alpha \ |\tilde W_O(\vec{\alpha}, \vec\theta)|\\
    ={}& \prod_{m\in S} \Big(\frac{2(1-\cos\theta_m)}{\pi}\int \dd^{2} \vec\alpha \ |\tilde W_{\ketbra{0}{0}}(\alpha_m, \theta_m)|\Big)\\
    &\cdot C_{\bar S} \int \dd^{2|\bar S|} \vec\alpha_{\bar S} \ |\tilde W_{O_{\bar S}}(\vec\alpha_{\bar S}, \vec\theta_{\bar S})|\\
    ={}& 2^{M-|\bar S|} C_{\bar S} Z_{O_{\bar S}}.
\end{split}
\end{equation}
Here, $C_{\bar S} = \prod_{m\in \bar S} [2(1-\cos\theta_m)/\pi]$. Since $O_{\bar S}$ is supported on at most $2N$ modes, which is independent of $M$ when $M > 2N$, the only $M$-dependence in $C_M Z_O$ comes from the $2^M$ factor. Unfortunately, this is still unfavorable, as it grows exponentially with mode number $M$.

To resolve this issue, consider $O_{\bar S}$, which is non-trivially supported on the modes contained in $\bar S$, rather than the full operator $O$ itself. We introduce the projection operator $P_S = \bigotimes_{m\in S} \ketbra{0}{0}_m$ and denote $\rho_{\bar S} = \Tr_S[\rho P_S]$. Here $\Tr_S[\bullet]$ means the partial trace over all modes with $m \in S$. We can see that
\begin{equation}
    \Tr[\rho O] = \Tr[\rho_{\bar S} O_{\bar S}],
\end{equation}
where $\rho_{\bar S}$ and $O_{\bar S}$ are wholly supported on modes in $\bar S$, which contains at most $2N$ elements. We can perform DEMESST sampling according to $O_{\bar S}$ as follows:
\begin{equation}
\begin{split}
    &\Tr[\rho_{\bar S} O_{\bar S}] \\
    ={}& C_{\bar S} Z_{O_{\bar S}} \int \dd^{2|\bar S|} \vec\alpha_{\bar S} \  p_{\text{D}}(\vec\alpha_{\bar S}) \ \re[e^{i\phi(\vec\alpha_{\bar S})}\tilde W_{\rho_{\bar S}}(\vec\alpha_{\bar S}, -\vec\theta_{\bar S})].
\end{split}
\end{equation}
In the experiment, we utilize the $\ket{f}$ level of the transmon to effectively restrict the cavity state to $\rho_{\bar S}$. From the measurement, we obtain binomial outcomes $A^{(k)}_j \in \{1, -1\}$ with expectation values $\re[e^{i\phi(\vec\alpha^{(k)}_{\bar S})}\tilde W_{\rho_{\bar S}}(\vec\alpha^{(k)}_{\bar S}, -\vec\theta_{\bar S})]$. More details about this experimental protocol is presented in Sec.~\ref{sec: exp-prtcl-anal}. Like before, we can use Hoeffding's inequality to estimate the sampling overhead. If we only measure once per sampling vector $\vec\alpha^{(k)}_{\bar S}$, we will have
\begin{equation}
    P\left(\left|\frac{C_{\bar S} Z_{O_{\bar S}}}{N_\spl} \sum_{k=1}^{N_\spl} A^{(k)} - \Tr[\rho_{\bar S} O_{\bar S}]\right| \geq \epsilon_2\right) \leq \delta_2
\end{equation}
when
\begin{equation} \label{eq: spl-oh-proj}
    N_\spl \geq \lceil \frac{2 C_{\bar S}^2 Z_{O_{\bar S}}^2}{\epsilon_2^2}\ln(2/\delta_2) \rceil.
\end{equation}
When $N$ is bounded, $C_{\bar S} Z_{O_{\bar S}}$ is independent of mode number $M$ when $M > 2N$. Therefore, $N_\spl$ in Eqn.~\eqref{eq: spl-oh-proj} scales as
\begin{equation}
    N_\spl \sim \mathcal{O}_M \Big( \frac{f(N)}{\epsilon_2^2}\ln(2/\delta_2) \Big),
\end{equation}
where $\mathcal{O}_M$ indicates that we focus only on the scaling over $M$ in the large $M$ limit, and $f(N) = 2 C_{\bar S}^2 Z_{O_{\bar S}}^2$ is a function that depend solely on $N$ and the specific form of $O$ from Eqn.~\eqref{eq: ops-basis}. We also introduce $f_{\max}(N)$ to represent the maximum value of $f(N)$ from those $\binom{M+N}{N}^2$ operators.

We can now consider our reconstructed density matrix $\hat \rho$. 
By performing expectation value estimation on the unknown state $\rho$ with all $\binom{M+N}{N}^2$ operators with form in Eqn.~\eqref{eq: ops-basis}, we can achieve
\begin{equation}
    P(B) \geq 1 - \binom{M+N}{N}^2 \delta_2
\end{equation}
with total sample number
\begin{equation}
    N_\text{tot} \sim \mathcal{O}_M \left[ \binom{M+N}{N}^2 \frac{f_{\max}(N)}{\epsilon_2^2}\ln(2/\delta_2) \right],
\end{equation}
where $B$ requires all the conditions below:
\begin{equation}
\begin{split}
    &|\bra{\vec n}(\hat \rho - \rho)\ket{\vec n}| \leq \epsilon_2, \\
    &|\re[\bra{\vec n}(\hat \rho - \rho)\ket{\vec n'}]| \leq \epsilon_2/\sqrt{2} \quad (\vec n \neq \vec n'), \\
    &|\im[\bra{\vec n}(\hat \rho - \rho)\ket{\vec n'}]| \leq \epsilon_2/\sqrt{2} \quad (\vec n \neq \vec n').
\end{split}
\end{equation}
If $B$ is satisfied, we will have the Frobenius norm distance between our reconstructed state and the true state satisfy
\begin{equation}
    ||\hat \rho - \rho||_F = \sqrt{\sum_{\vec n, \vec n'} |\bra{\vec n}(\hat \rho - \rho)\ket{\vec n'}|^2} \leq \binom{M+N}{N} \epsilon_2.
\end{equation}
Also, if $\rho$ is a pure state, we will have
\begin{equation}
\begin{split}
    |F(\hat \rho, \rho) - 1| &= |\Tr[(\hat \rho - \rho)\rho] \leq \epsilon_2 \cdot \sum_{\vec n, \vec n'} |\bra{n}\rho\ket{\vec n'}|\\
    & \leq \epsilon_2 \cdot \sqrt{\sum_{\vec n, \vec n'} |\bra{n}\rho\ket{\vec n'}|^2} \cdot \binom{M+N}{N} \\
    &= \epsilon_2 \cdot \sqrt{\Tr[\rho^2]} \cdot \binom{M+N}{N} = \binom{M+N}{N} \epsilon_2.
\end{split}
\end{equation}

In summary, by choosing $\epsilon_2 = \epsilon / \binom{M+N}{N}$ and $\delta_2 = \delta / \binom{M+N}{N}^2$, our sampling method will require the total sample number
\begin{equation}
\begin{split}
    &N_\text{tot} \sim \\
    &\mathcal{O}_M \left[ \binom{M+N}{N}^4 \frac{f_{\max}(N)}{\epsilon^2}\ln\left(2\binom{M+N}{N}^2/\delta\right) \right]
\end{split}
\end{equation}
to achieve
\begin{equation}
    P(||\hat \rho - \rho||_F \leq \epsilon) \geq 1 - \delta,
\end{equation}
even if we only perform a single measurement for each sampling instance. With the same amount of sampling, we can also achieve
\begin{equation}
    P(|F(\hat \rho, \rho) - 1| \leq \epsilon) \geq 1 - \delta
\end{equation}
when $\rho$ is a pure state.

The procedure for DEMESST is as follows: first, the Wigner function corresponding to a chosen basis operator is normalized to a probability distribution based on its absolute value. Then, displacement vectors are sampled from the resulting inverse cumulative distribution function (CDF) by randomly selecting a value between 0 and 1 to find the corresponding angular and radial values of the displacements (see supplementary information). This calculation is performed efficiently by utilizing Laguerre functions and their inverses. During measurement, each displacement vector will have the original sign of its Wigner function value preserved, so that if the Wigner function was negative at that point, the final measured value will be multiplied by -1. Observing a set of these will provide an estimate of the chosen density matrix element. Repeating this for multiple elements will thus produce the density matrix of the prepared state.

\section{DEMESST Experimental Protocol} \label{sec: exp-prtcl-anal}

For simplicity, in this section we assume that the cavity is initialized as a pure state $\ket{\psi}$. However, the same arguments will apply for a generic density operator $\rho$, which can always be decomposed as $\rho = \sum_i c_i \ketbra{\psi_i}{\psi_i}$ and understood as an ensemble average of a set of pure states $\{\ketbra{\psi_i}{\psi_i}\}$ with probability $c_i$.

First, we consider the generalized Wigner function of an $M$-mode state. We assume that the cavity-qubit state is initialized as $\ket{\psi}\ket{g}$. To perform the Wigner tomography measurement, we apply a short (large-bandwidth) drive to each cavity mode, then apply a large-bandwidth $\pi/2$ pulse on the qubit to begin the parity measurement. After these operations, the qubit part becomes $\exp(-i\frac{\pi}{4}\sigma_y) \ket{g} = \frac{\ket{g} + \ket{e}}{\sqrt{2}}$, and the cavity part becomes $\ket{\psi_D} = D(-\vec\alpha)\ket{\psi}$. Then, as part of the parity measurement, we wait for a time $t$. Due to the dispersive interaction Hamiltonian $H_{\text{int}} = \sum_m \chi_m a^\dag_m a_m \ketbra{e}{e}$, the cavity modes will be entangled with the qubit as $\frac{1}{\sqrt{2}} [\ket{\psi_D}\ket{g} + e^{-i\sum_m \theta_m a^\dag_m a_m}\ket{\psi_D}\ket{e}]$, where $\theta_m = \chi_m t$. In principle, we could choose any time $t$, as long as none of the $\theta_m$ are integer multiples of $2\pi$. However, in practice, we select $t$ to make each of the $\theta_m$ as close to $\pi$ modulo $2\pi$ as possible. This choice provides the maximum contrast and is closest to the ideal multimode parity operator. To complete our generalized parity measurement, we apply another $\pi/2$ pulse on the qubit, but with different phase from the initial one. Specifically, we consider the qubit rotation along $\vec r = -\sin\phi \ \vec e_x - \cos\phi \ \vec e_y$. By applying this $\exp(-i\frac{\pi}{4} \vec r\cdot \vec \sigma)$ operation, the final cavity-qubit entangled state $\ket{\Psi}$ will be
\begin{equation}
\begin{split}
    \ket{\Psi} ={}& \frac{\ket{\psi_D} + e^{i\phi} e^{-i\sum_m \theta_m a^\dag_m a_m}\ket{\psi_D}}{2} \ket{g}\\
    &+\frac{-e^{-i\phi}\ket{\psi_D} +  e^{-i\sum_m \theta_m a^\dag_m a_m}\ket{\psi_D}}{2} \ket{e}.
\end{split}
\end{equation}
Thus, when performing readout on the qubit, the final probability of achieving $\ket{g}$ will be
\begin{equation}
\begin{split}
    P_g &= \frac{1 + \re\{e^{i\phi}\Tr[D^\dag(\vec \alpha) \ketbra{\psi}{\psi} D(\vec \alpha) e^{-i\sum_m \theta_m a^\dag_m a_m}]\}}{2}\\
    &= \frac{1}{2} \{1 + \re[e^{i\phi}\tilde W_{\ketbra{\psi}{\psi}}(\vec \alpha, -\vec \theta)]\}.
\end{split}
\end{equation}
Therefore, if we record $A = 1$ upon measuring $\ket{g}$ and $A = -1$ otherwise, the expectation value of $A$ will be exactly $\re[e^{i\phi}\tilde W_\rho(\vec \alpha, -\vec \theta)]$. This derivation applies for any $\phi$, but as mentioned before, the choice of $\phi$ depends on the operator $O$ and the sampling vector $\vec \alpha$.

We must modify the experimental protocol above to measure the generalized Wigner function for the projected state $\rho_{\bar S}$, which is defined in Sec.~\ref{sec: dm-recstr}. In particular, we utilize the second excited state $\ket{f}$ of the transmon to implement subsystem tomography and measure the Wigner values for only the projected states $\rho_{\bar S}$, which is similar to the idea used in~\cite[S][]{gertler_experimental_2023}. 
Our W states are generated using multimode photon blockade as described in~\cite[S][]{chakram_multimode_2022} to ensure that our maximum total photon number is $N = 1$. Consequently, for the density matrix reconstruction, the Hilbert space will be spanned by $\{\ket{\vec n}| \sum_{m=1}^M n_m \leq 1\}$. In this case, the projected operator $O_{\bar S}$ introduced in Eqn.~\eqref{eq: op-decomps} will be supported on at most $2$ modes. 
Because of the different dispersive couplings between the qubit and distinct cavity modes (STable~\ref{parameters_table}), we can selectively target each of the modes with sufficiently narrow-bandwidth qubit pulses to help perform the necessary projections.
Therefore, before the parity measurement, we first apply several long (narrow-bandwidth) qubit $\pi$ pulses with frequencies $\omega_q + \chi_m$ for $m\in S$, such that the qubit that coupled with the $(I - P_S)\ket{\psi}$ component of the multimode cavity state will transfer from $\ket{g}$ to $\ket{e}$, while the component that coupled with $P_S\ket{\psi}$ will stay in $\ket{g}$. Then we give the transmon a short $\pi$ pulse on the $\ket{e}-\ket{f}$ transition. After those steps, the cavity-transmon state becomes
\begin{equation}
    \ket{\Psi} = P_S\ket{\psi}\ket{g} + (I - P_S)\ket{\psi}\ket{f}.
\end{equation}
Finally, we can use the procedure that we described before when focusing on the Wigner value measurement of a generic $M$-mode state $\rho$. We only need to drive those modes with index $m\in \bar S$ such that those modes are displaced by $D(-\vec \alpha_{\bar S})$. The probability to measure $\ket{g}$ from the final qubit readout is
\begin{equation}
    P_{g, \phi} = \frac{\Tr[\rho_{\bar S}] + \re[e^{i\phi} \tilde W_{\rho_{\bar S}}(\vec \alpha_{\bar S}, -\vec \theta_{\bar S})]}{2},
    \label{eqn: original_phase_exp_protocol}
\end{equation}
where $\rho_{\bar S} = \Tr_S[P_S \ketbra{\psi}{\psi}]$. In practice, the result above is unaffected by the order in which we perform the qubit $\pi$ pulses and cavity displacements, assuming that we add additional $\pi$ pulses to target the displaced state. We found this order to work better in the experiment, despite the need for a larger comb of $\pi$ pulse frequencies.

Now, if we proceed similarly to before and assign $A = 1$ for $\ket{g}$ and $A = -1$ otherwise, the expectation value of $A$ will not give us our desired result. To solve this issue, we utilize the freedom we have in choosing the phase of the second qubit $\pi/2$ pulse in the parity measurement. If we choose the second qubit $\pi/2$ rotation to be along $-\vec r$ instead of $\vec r$ (or equivalently choose $(\phi+\pi)$ instead of $\phi$, then the probability of measuring $\ket{g}$ will be
\begin{equation}
    P_{g, (\phi+\pi)} = \frac{\Tr[\rho_{\bar S}] - \re[e^{i\phi} \tilde W_{\rho_{\bar S}}(\vec \alpha_{\bar S}, -\vec \theta_{\bar S})]}{2}.
\end{equation}
Therefore, comparing with Eqn.~\eqref{eqn: original_phase_exp_protocol}, one solution to recovering our desired quantity is to first generate a random binomial number $s$. There is a $50\%$ probability that $s=1$ and $50\%$ probability that $s=-1$. If we get $s=1$, we choose the second qubit $\pi/2$ rotation to be along $\vec r$, and otherwise choose $-\vec r$ instead. In both cases, we assign $A=1$ if the qubit measurement outcome is $\ket{g}$, and assign $A=-1$ if it is not $\ket{g}$. The expectation value of $sA$ will be exactly $ \re[e^{i\phi} \tilde W_{\rho_{\bar S}}(\vec \alpha_{\bar S}, -\vec \theta_{\bar S})]$. An advantage of this procedure is that it will work even when it is difficult to distinguish $\ket{e}$ and $\ket{f}$ levels in qubit readout, as long as we can distinguish $\ket{g}$ outcomes from others. The same applies for other permutations of these three states, as long as the experimental protocol is adjusted accordingly. In the actual experiment, we did not use this trick of random number $s$ generation, since we can perform more than a single measurement per sampling vector $\vec \alpha_{\bar S}$. Instead, we repeated the experiment $10$ times for rotation of the second $\pi/2$ along $\vec r$ and $10$ times along $-\vec r$. Finally, we subtracted the averaged probability of measuring $\ket{g}$ between the two cases to obtain an estimate for $\re[e^{i\phi} \tilde W_{\rho_{\bar S}}(\vec \alpha_{\bar S}, -\vec \theta_{\bar S})]$.

\section{Final Reconstructed Density Matrices}
\begin{figure}
  % \begin{center}
    \includegraphics[width=0.48\textwidth]{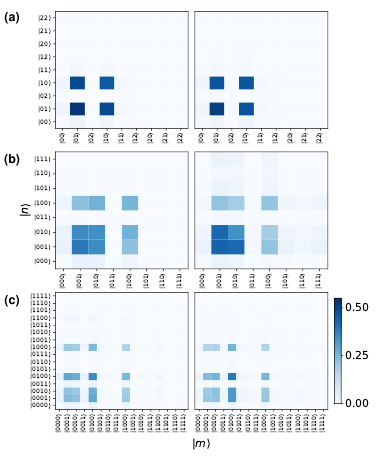}
    \caption{Absolute value of the final density matrices determined using the DEMESST tomography sampling method (left column) and the OLI method (right column) for W states of (a) 2, (b) 3, and (c) 4 modes. The results for DEMESST and OLI are in good agreement.}
  	\label{final_dms}
  % \end{center}
\end{figure} 

Using the DEMESST and OLI tomography methods, we can reconstruct the final density matrices of our prepared W states. These matrices are final in the sense that they are the results obtained from the entire set of measurements performed in the experiment, i.e. 100 averages for each distinct displacement, so that the total measurement number is 10 times the maximum measurement number shown in Fig.~3 of the main text. Given a fixed total measurement number that equals the product of a number of distinct sets of cavity displacements times the number of averages that each displacement is repeated, we would gain the most information from maximizing the number of distinct displacements and measuring each a single time. We choose instead to average each measurement 10 times due to our imperfect readout fidelity, to minimize our average number but still be able to obtain accurate measurement results. We then repeat this process 10 times to obtain statistics, resulting in a total of 100 averages for each distinct displacement. This choice lets us balance this theoretical maximal information of singleshot measurements with our measurement errors.

The final reconstructed density matrices for 2, 3, and 4 modes are shown in SFig.~\ref{final_dms}. We plot the absolute values of the density matrix elements so that we have a single matrix grid for each combination of tomography method and mode number. We can see that the two methods are in good agreement, with the largest visible deviation being in the 3-mode case for Fock basis elements with nonzero population in the second (middle) cavity mode. Nevertheless, the distances between the two final matrices as determined by the two methods is still low, and is 0.05 for the 2-mode case, 0.22 for the 3-mode case, and 0.30 for the 4-mode case. These distances are all below the corresponding minimum distances at the maximum total measurement number presented in the main text, and so this difference should not have significantly affected those results. 

Besides some slight deviations in the measured populations of individual density matrix elements, the remaining distance between the final reconstructed matrices can be explained by small differences in the fit phase angles of the W states. For the 2-mode W state with form $\ket{W_2} = (\ket{10} + e^{i \phi} \ket{01}) / \sqrt{2}$, using the DEMESST method we obtain a fit $\phi_D = 0.04$, and using the OLI method we measure $\phi_{O} = 0.03$. In the 3-mode case, for W state with phase angles defined as $\ket{W_3} = (\ket{100} + e^{i \phi_1} \ket{010} + e^{i \phi_2} \ket{001}) / \sqrt{3}$, we measure $\phi_{1D} = -0.19$ and $\phi_{2D} = 1.57$, while $\phi_{1O} = -0.12$ and $\phi_{2O} = 1.57$. Similarly to what we see in the populations, the deviation is primarily in the middle mode. Finally, for 4 modes and W state $\ket{W_4} = (\ket{1000} + e^{i \phi_1} \ket{0100} + e^{i \phi_2} \ket{0010} + e^{i \phi_3} \ket{0001}) / \sqrt{4}$, we find $\phi_{1D} = -1.36, \phi_{2D} = -2.90$, and $\phi_{3D} = 0.60$, and $\phi_{1O} = -1.38, \phi_{2O} = -3.02$, and $\phi_{3O} = 0.63$. These angles are obtained by discretely sweeping the $\phi_j$ values over the full $2\pi$ range for each of the modes and determining which set of $\phi_j$ gives the largest fidelity when compared to an ideal W state with those phases.

\section{Infidelity and Matrix Distance Simulations}
\begin{figure}
  % \begin{center}
    \includegraphics[width=0.4\textwidth]{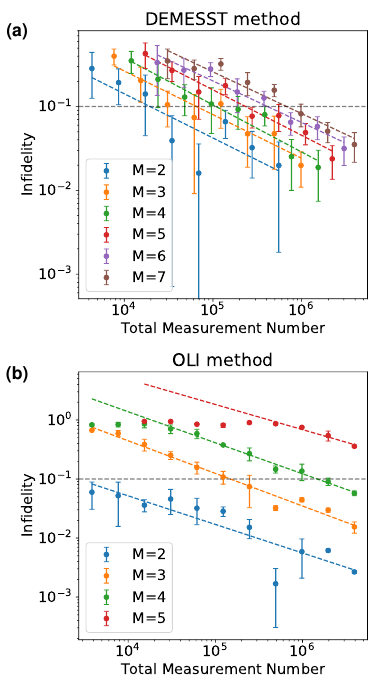}
    \caption{Simulated infidelity vs. total measurement number for W states of different mode numbers for the two tomography methods, (a) DEMESST and (b) OLI. The infidelity is computed as the fidelity difference between the reconstructed state at a given total measurement number vs. an ideal $M$-mode W state. Error bars are obtained from repeating the simulation multiple times while including readout bit flip errors. The infidelities decrease to lower values more quickly for the DEMESST approach, especially for larger mode numbers. The dashed horizontal lines indicates an infidelity of 0.1 (90\% fidelity).}
  	\label{fid_sims}
  % \end{center}
\end{figure}

In this section, we present simulations of the infidelity and Frobenius norm matrix distance vs. point number for our two tomography methods, DEMESST and OLI, following the same procedure described in the main text. Wigner tomography measurements are sampled assuming perfect state preparation, and the infidelities are computed with respect to an ideal $M$-mode W state, and the results are shown in SFig.~\ref{fid_sims}. Error bars are obtained by repeating the simulations while modeling bit flip readout errors. The resulting infidelity vs. total measurement number for each $M$ is fit to a power law, and the intersection with 0.1 (90\% fidelity) is used to generate the values plotted in Fig.~1(a) in the main text. We can see that the OLI method requires fewer measurements than DEMESST for 2 modes, but DEMESST has a lower sampling requirement for 3 or more modes. This effect becomes increasingly apparent for larger $M$. DEMESST scales polynomially with $M$, while OLI scales exponentially with $M$.

\begin{figure}
  % \begin{center}
    \includegraphics[width=0.4\textwidth]{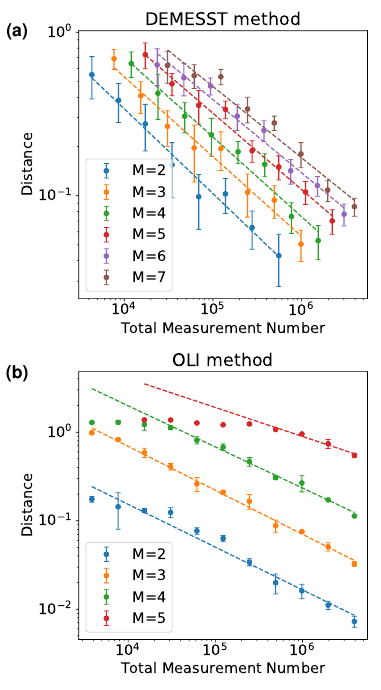}
    \caption{Simulated matrix distance vs. total measurement number for W states of different mode numbers for the two tomography methods, (a) DEMESST and (b) OLI. The distance is computed as the Frobenius norm between the reconstructed state at a given measurement number and a final simulated W state with state preparation errors from photon blockade and decoherences. Error bars are obtained from repeating the simulation multiple times while including readout bit flip errors. The distances decrease to lower values more quickly for the DEMESST approach, especially for larger mode numbers.}
  	\label{distance_sims}
  % \end{center}
\end{figure}

To compute the Frobenius norm matrix distances, we obtain the density matrix of an imperfect W state prepared by photon blockade, with errors from transmon and cavity decoherence and leakage through the blockade. We then simulate Wigner tomography measurements on that state for the total measurement numbers and using the same cavity displacement sampling points as used in our experiment, then reconstruct the density matrix while including readout and bit flip error. These simulated results are shown in SFig.~\ref{distance_sims}, with error bars obtained from repeating the simulations with the readout errors. As expected, the DEMESST method performs increasingly more efficiently as the mode number increases. 

Comparing to the experimental data presented in the main text, the matrix distance results are similar, albeit with some differences. For example, we can see that the measurement number at which the simulated distance reaches roughly 0.1 for 2 modes is slightly less than $10^5$, while we observe a distance slightly above 0.1 at that point number in our data. For the 3-mode case, in the experiment we observe a distance of roughly 0.3 at $10^5$ measurements for DEMESST and 0.4 for OLI, which is close to the simulated distances of roughly 0.25 and 0.3, respectively. For the 4-mode case, we measure a distance of 0.4 at roughly $2\times 10^5$ measurements for DEMESST, compared to a simulated distance of roughly 0.2, and a distance of roughly 1.0 vs 0.8 for OLI at that measurement number. We attribute the discrepancies to fluctuations over time in the readout error that may affect the accuracy of the simulation. This effect is particularly pronounced for the 4-mode case, where more measurements are required.

\section{$W^2$ State Reconstruction}
\begin{figure}
  % \begin{center}
    \includegraphics[width=0.483\textwidth]{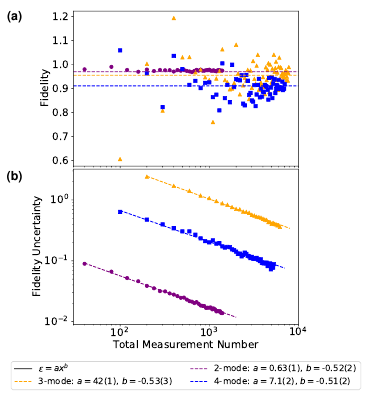}
    \caption{Experimental results for the $W^2$ tomography sampling method. (a) State reconstruction fidelities for 2- (purple circles), 3- (orange triangles), and 4-mode (blue squares) W states. Horizontal lines indicate the fidelities obtained from the OLI method, which are consistent with the $W^2$. (b) Error magnitudes for the fidelities shown in (a). All results follow a roughly 1/$\sqrt{x}$ relationship vs. total measurement number, as expected.}
  	\label{SFig_W2_data}
  % \end{center}
\end{figure}
Another Wigner tomography sampling method that we implement is the $W^2$ method, which was first introduced in~\cite[S][]{da_silva_practical_2011} and that we briefly discussed in Sec.~\ref{sec: wig-sampling}. In this approach, sets of coherent cavity displacements $\alpha_i$ are chosen using rejection sampling. This approach computes the overlap between a prepared state and a desired target state. For $M$ modes, a cutoff $c$ and a displacement vector $(\beta_1, ..., \beta_M)$ is randomly sampled from a uniform distribution between 0 and a maximum value of $|\mathcal{W}(\alpha_1, ..., \alpha_M)|^2 \prod_{i=1}^M |\alpha_i|$, where $\mathcal{W}$ corresponds to the target state. If $|\mathcal{W}(\beta_1, ..., \beta_M)|^2 \prod_{i=1}^M |\beta_i| > c$, the vector is kept. This ensures that we measure cavity displacements that provide the most information about the state, while also avoiding displacements with large magnitude or Wigner values near zero, which are more susceptible to experimental errors. After measuring a set of $n$ of these vectors, the final overlap fidelity is computed as $\frac{1}{n} \sum_{i=1}^n \mathcal{W}_{\text{exp}}(\overrightarrow{\alpha_i}) / \mathcal{W}_{\text{ideal}}(\overrightarrow{\alpha_i})$. This approach allows for direct fidelity estimation of a prepared state with an ideal state.
In particular, the $W^2$ method can be used in a similar manner to the DEMESST, where the fidelity estimation is performed with respect to multimode Fock state basis elements. Repeating for multiple elements can thus provide a reconstructed density matrix. 

Experimentally, we use the $W^2$ method as an additional check on our prepared W states. We set the target state to be the multimode W state with $\phi$'s determined from the DEMESST and OLI methods. The $W^2$ measurements then provide a direct fidelity estimation of the prepared state with the expected target. Since the sampling uses these angles, we present the $W^2$ results independently from the DEMESST and OLI, which do not utilize that information. The results are shown in SFig.~\ref{SFig_W2_data}. The fidelities for the maximum provided observation number are $0.972 \pm 0.013$, $0.95\pm0.35$, and $0.90\pm0.08$ for the 2-, 3-, and 4-mode W states respectively. These averages are consistent with the results of the previous DEMESST and OLI methods, with the OLI fidelities indicated by the horizontal lines in SFig.~\ref{SFig_W2_data}(a), and the data converges quickly to the expected fidelity obtained from those two approaches, although with large uncertainties, as shown in SFig.~\ref{SFig_W2_data}(b). One reason for these errors is the relatively low total measurement number compared with the other methods. However, an odd behavior is that the 3-mode data has much greater uncertainties than even the 4-mode case, when we would expect the uncertainties to increase monotonically with mode number. Some possible explanations for this behavior could be a particularly low readout fidelity during data collection or fluctuations in drive strength during the measurement sequence that modify the effective Wigner operator differently for distinct sets of cavity mode displacements. This could also be caused by the choice in cutoff, as derived in Sec.~\ref{sec: wig-sampling}.
All the uncertainties have the expected $1/\sqrt{x}$ with total observation number.

% \begin{center}
%     \textbf{\small REFERENCES}
% \end{center}
% \printbibliography[heading=none]
% \end{refsection}

% \end{bibunit}

\end{document}